\documentclass[useAMS,usenatbib]{mnras}

\usepackage{graphicx}	% Including figure files
\usepackage{subcaption}
\usepackage{floatrow}
\usepackage{amsmath}	% Advanced maths commands
\usepackage{amssymb}	% Extra maths symbols
\usepackage{multicol}        % Multi-column entries in tables
\usepackage{bm}		% Bold maths symbols, including upright Greek

\newcommand{\vect}[1]{\boldsymbol{#1}}

\newcommand{\dif}{\mathrm{d}}

\newcommand{\Myr}{\,\mathrm{Myr}}
\newcommand{\Gyr}{\,\mathrm{Gyr}}
\newcommand{\AU}{\,\mathrm{au}}

\newcommand{\RE}{R_{\oplus}}
\newcommand{\MJ}{M_{\rm J}}
\newcommand{\RJ}{R_{\rm J}}
\newcommand{\MSol}{M_{\odot}}

\newcommand{\tLK}{t_{\rm LK}}

\newcommand{\ehat}{\vect{\hat{e}}}
\newcommand{\Lhat}{\vect{\hat{L}}}
\newcommand{\zhat}{\vect{\hat{z}}}

\title[LK migration of WD\,1856\,b]{Enhanced Lidov--Kozai migration and the formation of the transiting giant planet WD\,1856+534\,b}

\author[O'Connor, Liu \& Lai]
{Christopher E. O'Connor$^{1}$\thanks{E-mail: coconnor@astro.cornell.edu}, Bin Liu$^{1}$, and Dong Lai$^{1,2}$ \\
$^{1}$Cornell Center for Astrophysics and Planetary Science, Department of Astronomy, Cornell University, Ithaca, NY 14853, U.S.A. \\
$^{2}$Tsung-Dao Lee Institute, Shanghai Jiao Tong University, Shanghai 200240, China
}

\begin{document}

\date{Accepted 2020 November 25. Received 2020 November 22; in original form 2020 October 7.}

\pagerange{\pageref{firstpage}--\pageref{lastpage}} \pubyear{2020}

\maketitle

\label{firstpage}

\begin{abstract}
    We investigate the possible origin of the transiting giant planet WD\,1856+534\,b, the first strong exoplanet candidate orbiting a white dwarf, through high-eccentricity migration (HEM) driven by the Lidov--Kozai (LK) effect. The host system's overall architecture is an hierarchical quadruple in the `2+2' configuration, owing to the presence of a tertiary companion system of two M-dwarfs. We show that a secular inclination resonance in 2+2 systems can significantly broaden the LK window for extreme eccentricity excitation ($e \gtrsim 0.999$), allowing the giant planet to migrate for a wide range of initial orbital inclinations. Octupole effects can also contribute to the broadening of this `extreme' LK window. By requiring that perturbations from the companion stars be able to overcome short-range forces and excite the planet's eccentricity to $e \simeq 1$, we obtain an absolute limit of $a_{1} \gtrsim 8 \AU \, (a_{3} / 1500 \AU)^{6/7}$ for the planet's semi-major axis just before migration (where $a_{3}$ is the semi-major axis of the `outer' orbit). We suggest that, to achieve a wide LK window through the 2+2 resonance, WD\,1856\,b likely migrated from $30 \AU \lesssim a_{1} \lesssim 60 \AU$, corresponding to $\sim 10$--$20 \AU$ during the host's main-sequence phase. We discuss possible difficulties of all flavours of HEM affecting the occurrence rate of short-period giant planets around white dwarfs.
\end{abstract}


\section{Introduction}

The recent discovery of WD\,1856+534\,b (hereafter WD\,1856\,b), a candidate giant planet transiting its white-dwarf (WD) host with a period of 1.4 days \citep{Vanderburg+2020}, is a major milestone in the emerging field of WD planetary science. The existence of remnant planetary systems around WDs has been inferred from observations of a disintegrating planetesimal \citep[e.g.,][]{Vanderburg+2015}, gas accretion from an evaporating ice giant \citep{Gansicke+2019}, infrared emission from debris discs \citep[e.g.,][]{FJZ2009}, and atmospheric pollution \citep{Zuckerman+2003,Zuckerman+2010,KGF2014}. These phenomena are thought to be driven by tidal disruption events \citep{Jura2003}, perhaps triggered dynamically by distant planets \citep{DS2002,DWS2012,FH2014,PML2017,Mustill+2018} or stellar binary partners \citep{BV2015,HPZ2016b,PM2017,SNZ2017}.

Planets orbiting within a few AU of their host stars during the main sequence are likely to be engulfed and destroyed when the star ascends the asymptotic giant branch (AGB) on its way to becoming a WD \citep{VL2007,MV2012}. The dynamical history of WD\,1856b may then be similar to the hypothesized origin of hot Jupiters around main-sequence stars by high-eccentricity migration (HEM; reviewed by \citealt{DJ2018}). During HEM, a planet's orbital eccentricity $e$ is excited to an extreme value ($1 - e \lesssim 10^{-2}$) via dynamical interactions. The planet then experiences strong tidal dissipation near pericenter, which shrinks and circularizes its orbit. Possible mechanisms for driving HEM are diverse, including secular interactions with other giant planets or binary companions \citep[e.g.,][]{WM2003,FT2007,NFR2012,WL2011,Petrovich2015,Petrovich2015b,ASL2016,Hamers+2017,VLA2019,TLV2019}. Similar mechanisms are plausible in a WD's planetary system. Importantly, a successful model for the origin of WD\,1856\,b must be able to delay migration until the host star becomes a WD, since a planet migrating sooner than this would presumably have been destroyed.

For WD\,1856\,b, HEM by way of the Lidov--Kozai (LK) effect \citep{vonZeipel1910,Lidov1962,Kozai1962} is an inviting hypothesis because its host star belongs to a hierarchical triple system: \citet{Vanderburg+2020} identified two bound M-dwarf companions with a projected separation $\sim 1000 \AU$ from the primary WD and $\sim 50 \AU$ apart; we summarize the existing constraints on the system's properties in Table \ref{tab:properties}. Including the planet, the system's overall architecture is a hierarchical quadruple system with a `2+2' configuration (see Figure \ref{fig:cartoon}). The dynamics of such systems resembles the standard LK effect in some respects but allows for the excitation of extreme eccentricities from a wider range of initial conditions \citep{Pejcha+2013,HL2017,FTH2018}.

In this paper, we demonstrate the possible origin of the WD\,1856\,b through HEM by way of such ``enhanced'' secular interactions with the WD's stellar companions. We show that this could have occurred if the planet's semi-major axis prior to migration was $\sim 30$--$60 \AU$, corresponding to $\sim 10$--$20 \AU$ during the host's main sequence. In Section \ref{s:Dyn}, we describe the secular dynamics of a 2+2 hierarchical system and the conditions for HEM in such systems. In Section \ref{s:HEM_ex}, we apply our dynamical model to HEM in the WD\,1856 system. We examine the system's early dynamical history in Section \ref{s:Disc:preWD}, discuss potential issues of HEM affecting the occurrence rate of giant planets around WDs in Section \ref{s:Disc:Cav}, and summarize in Section \ref{s:Sum}.

\section{Secular Dynamics of a 2+2 Hierarchical System} \label{s:Dyn}

\begin{figure}
    \centering
    \includegraphics[width=\columnwidth]{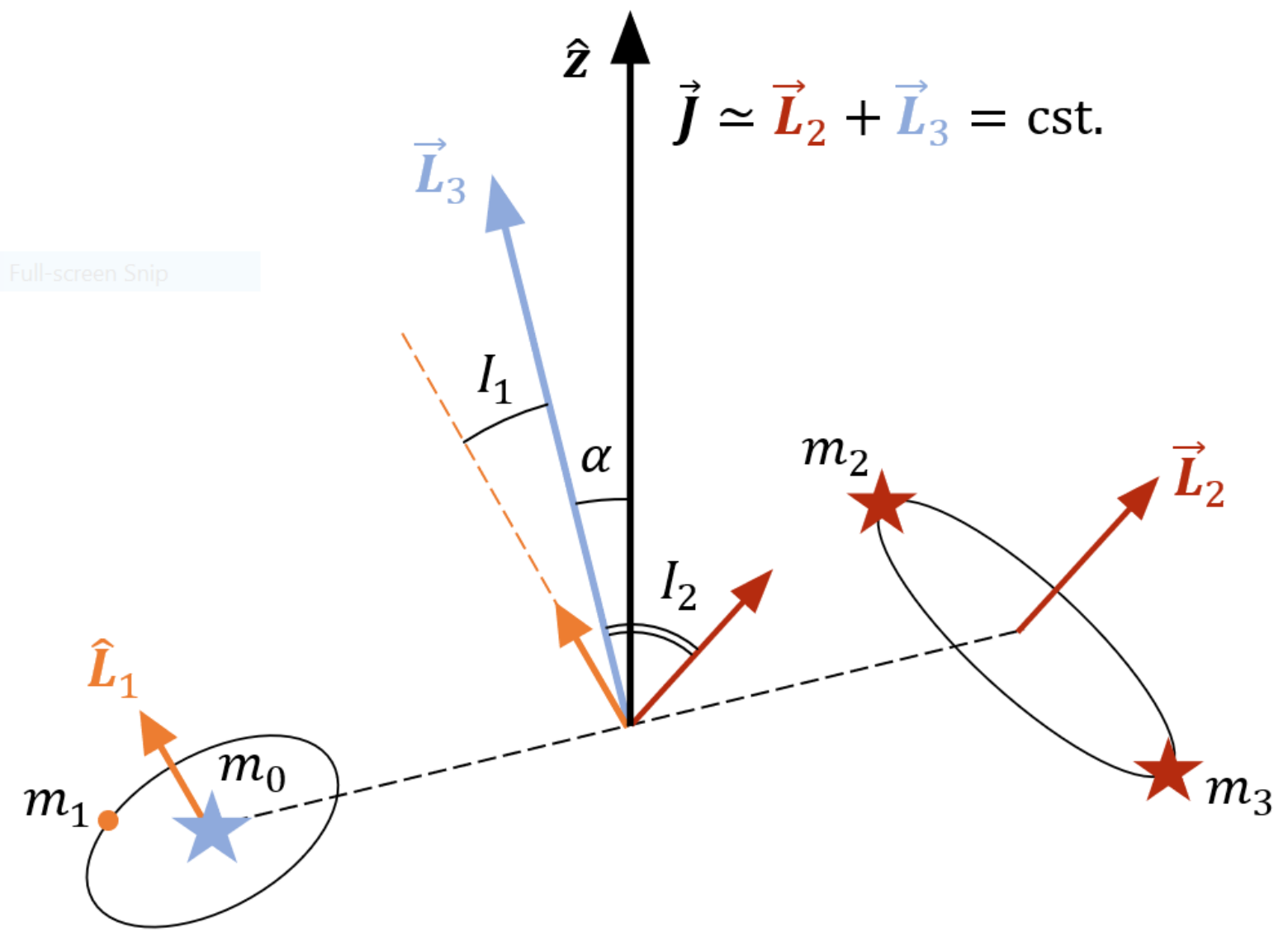}
    \caption{Schematic cartoon of a hierarchical 2+2 system with $m_{1} \ll m_{0}, m_{2}, m_{3}$.}
    \label{fig:cartoon}
\end{figure}

\begin{table}
    \centering
    \begin{tabular}{lll}
        \hline
        Quantity & Symbol & Value \\
        \hline
        WD mass & $m_{0}$ & $0.6 \MSol$ (assumed) \\
        Planetary mass & $m_{1}$ & $1.0 \MJ$ (assumed) \\
        Planetary radius & $R_{1}$ & $(0.93 \pm 0.09) \RJ$ \\
        Companion masses & $m_{2}$ & $(0.346 \pm 0.027) \MSol$ \\
         & $m_{3}$ & $(0.331 \pm 0.024) \MSol$ \\
        Semi-major axes & $a_{1,{\rm obs}}$ & $\approx 0.02 \AU$ \\
         & $a_{2}$ & $58^{+54}_{-16} \AU$ \\
         & $a_{3}$ & $1500^{+700}_{-240} \AU$ \\
        Eccentricities & $e_{1,{\rm obs}}$ & 0 (assumed) \\
         & $e_{2}$ & $<0.63$ \\
         & $e_{3}$ & $0.30^{+0.19}_{-0.10}$ \\ \hline
    \end{tabular}
    \caption{Properties of the WD\,1856 system. All values are as reported by \citet{Vanderburg+2020} except for the WD and planetary masses, which have not been robustly measured (see Section \ref{s:Disc:Cav:props}).}
    \label{tab:properties}
\end{table}

Figure \ref{fig:cartoon} shows a schematic depiction of a 2+2 hierarchical quadruple system. The Hamiltonian describing the secular evolution of such systems has been calculated up to the quadrupole order of approximation by \citet{Hamers+2015}, \citet{HPZ2016a}, and \citet{FTH2018}. \citet{HL2017} studied the limit where one body is a test particle and identified the key dynamical mechanism for an enhanced LK effect. We adopt their setup in this paper.

A 2+2 system can be described by three quasi-Keplerian orbits: two ``inner'' binary systems (labelled 1 and 2) orbit their relative barycentres, which in turn follow an ``outer'' orbit (3) around the system's total centre of mass. We label the component masses $m_{0}$ and $m_{1}$ for orbit 1 and $m_{2}$ and $m_{3}$ for 2. For each orbit, we denote the eccentricity vector $\vect{e}_{k} = e_{k} \ehat_{k}$ and angular momentum vector $\vect{L}_{k} = \mu_{k} [ G M_{k} a_{k} (1 - e_{k}^{2}) ]^{1/2} \Lhat_{k}$, where $\mu_{k}$, $M_{k}$, and $a_{k}$ are respectively the reduced mass, total mass, and semi-major axis. We also define the dimensionless angular momentum $\vect{j}_{k} = (1 - e_{k}^{2})^{1/2} \Lhat_{k}$ and the total angular momentum $\vect{J} = \vect{L}_{1} + \vect{L}_{2} + \vect{L}_{3} \equiv J \zhat$. Finally, we define $\cos I_{1} = \Lhat_{1} \cdot \Lhat_{3}$, $\cos I_{2} = \Lhat_{2} \cdot \Lhat_{3}$, and $\cos\alpha = \Lhat_{3} \cdot \zhat$ (see Fig.~\ref{fig:cartoon}).

We consider a hierarchical system with $m_{1} \ll m_{0,2,3}$ such that $L_{1} \ll L_{2}$, $\vect{J} \simeq \vect{L}_{2} + \vect{L}_{3} = {\rm const.}$, and $L_{2} \sin I_{2} \simeq L_{3} \sin \alpha$. For simplicity, we assume that $\vect{e}_{2} = \mathbf{0}$ and that the mutual inclination of orbits 2 and 3 is $I_{2} < 39 \fdg 2$ in order to suppress the LK effect for orbit 2. The latter two assumptions may not hold for general 2+2 systems, but they are appropriate for the WD\,1856 system.\footnote{For general $e_{2}$ and $I_{2}$, the precession rate of $\vect{L}_{3}$ around $\vect{J}$ and the angle $\alpha$ between $\vect{L}_{3}$ and $\vect{J}$ both vary due to eccentricity oscillations. All these variations occur on a similar time-scale per Eq.~(\ref{eq:Om3z}). Thus we do not expect these complications to appreciably change the dynamics.}

Under these assumptions, the secular evolution is completely described by the following equations:
\begin{align}
    \frac{\dif \vect{j}_{1}}{\dif t} &= \frac{3}{4 \tLK} \left[ \left( \vect{j}_{1} \cdot \Lhat_{3} \right) \left( \vect{j}_{1} \times \Lhat_{3} \right) \right. \nonumber \\ & \left. \hspace{1.5cm} -\, 5 \left( \vect{e}_{1} \cdot \Lhat_{3} \right) \left( \vect{e}_{1} \times \Lhat_{3} \right) \right], \label{eq:vecteqs_dj1dt} \\
    \frac{\dif \vect{e}_{1}}{\dif t} &= \frac{3}{4 \tLK} \left[ \left( \vect{j}_{1} \cdot \Lhat_{3} \right) \left( \vect{e}_{1} \times \Lhat_{3} \right) + 2 \left( \vect{j}_{1} \times \vect{e}_{1} \right) \right. \nonumber \\ & \left. \hspace{1.5cm} -\, 5 \left( \vect{e}_{1} \cdot \Lhat_{3} \right) \left( \vect{j}_{1} \times \Lhat_{3} \right) \right], \label{eq:vecteqs_de1dt} \\
    \frac{\dif \Lhat_{3}}{\dif t} &= - \frac{\beta}{\tLK} \left( \zhat \times \Lhat_{3} \right). \label{eq:vecteqs_dL3dt}
\end{align}
Above we have defined the LK time-scale for orbit 1 as
\begin{equation} \label{eq:tLK_def}
    \frac{1}{\tLK} = \frac{m_{23}}{m_{0}} \left( \frac{a_{1}}{a_{3,{\rm eff}}} \right)^{3} n_{1},
\end{equation}
where $m_{23} = m_{2} + m_{3}$, $a_{3,{\rm eff}} = a_{3} (1 - e_{3}^{2})^{1/2}$, and $n_{1} = ( G m_{0} / a_{1}^{3} )^{1/2}$. We have also defined a dimensionless quantity
\begin{equation} \label{eq:beta_def}
    \beta = \frac{3}{4} \left( \frac{m_{0}}{m_{23}} \frac{a_{2}}{a_{1}} \right)^{3/2} \frac{J}{L_{3}} \cos I_{2},
\end{equation}
which is simply the precession rate of $\vect{L}_{3}$ about $\zhat$,
\begin{equation} \label{eq:Om3z}
    \Omega_{3z} = \frac{3}{4} \frac{m_{0}}{m_{23}} \left( \frac{a_{2}}{a_{3,{\rm eff}}} \right)^{3} \frac{J}{L_{2}} n_{2},
\end{equation}
in units of $\tLK^{-1}$, i.e. $\beta = \Omega_{3z} \tLK$.

\begin{figure*}
    \centering
    \includegraphics[width=0.9\textwidth]{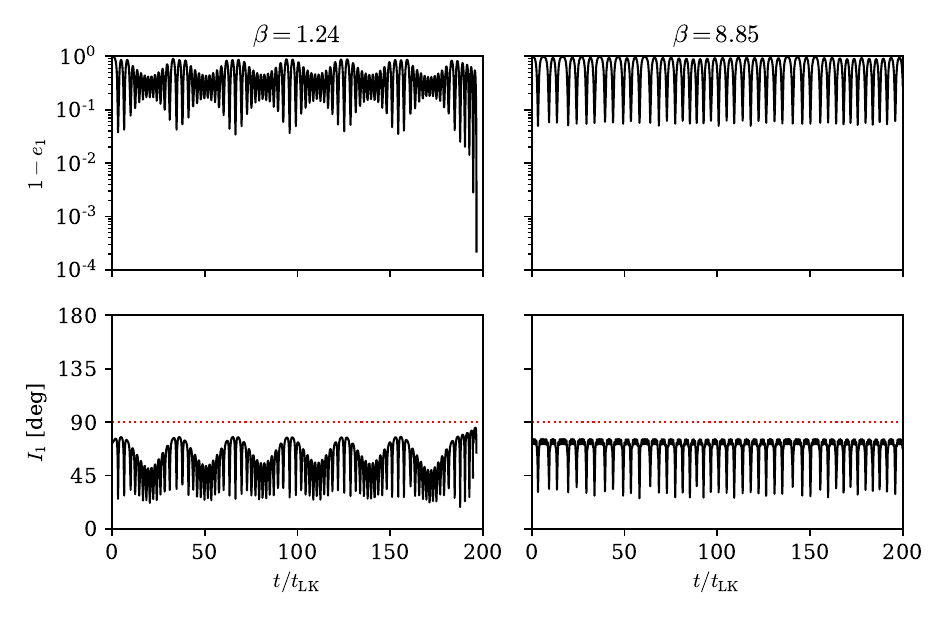}
    \caption{Eccentricity (as $1-e_{1}$; upper panels) and inclination (lower) of the test particle as a function of time for trajectories in the resonant regime ($\beta \sim 1$, left panels) and modified LK ($\beta \gg 1$, right) regime. In these examples, we use identical initial conditions $I_{1,0} = 75^{\circ}$, $\Omega_{1,0} = 135^{\circ}$, $e_{1,0} = 0.01$, $\omega_{1,0} = 344^{\circ}$.}
    \label{fig:compare_regimes}
\end{figure*}

The qualitative nature of the dynamics is determined by the value of $\beta$ \citep{HL2017}. There are three regimes:
\begin{enumerate}
    \item[(i)] When $\beta \ll 1$, $\Lhat_{1}$ precesses around $\Lhat_{3}$ more rapidly than $\Lhat_{3}$ around $\zhat$. The system's dynamical evolution approaches that of a `2+1' hierarchical triple in the standard LK problem. If the initial mutual inclination of orbits 1 and 3 satisfies $\cos I_{1,0} < \sqrt{3/5}$, then the eccentricity of $m_{1}$ can be excited according to
    \begin{equation} \label{eq:std_LK_law}
        e_{1,{\rm max}} = \sqrt{1 - \frac{5}{3} \cos^{2} (I_{1,0})}.
    \end{equation}
    We call this the ``quasi-LK'' regime.
    \item[(ii)] When $\beta \gg 1$, $\Lhat_{1}$ precesses around $\Lhat_{3}$ slowly compared to $\Lhat_{3}$ around $\zhat$; one can then average over the precession of $\Lhat_{3}$, so that $\Lhat_{1}$ effectively precesses around $\zhat$. The dynamics of $m_{1}$ resembles the LK problem with a modified conservation law:
    \begin{equation} \label{eq:mod_LK_law}
        e_{1,{\rm max}} = \sqrt{1 - \frac{5}{3} \cos^{2} (\theta_{1,0})},
    \end{equation}
    where $\theta_{1,0}$ is the initial angle between $\Lhat_{1}$ and $\zhat$. We call this the ``modified LK'' regime.
    \item[(iii)] When $\beta \sim 1$, $\Lhat_{1}$ precesses around $\Lhat_{3}$ at roughly the same rate as $\Lhat_{3}$ around $\zhat$. This secular inclination resonance drives chaotic evolution of orbit 1, featuring extreme eccentricities for a broader range of initial inclinations -- an ``enhanced LK window.'' We call this the ``resonant'' regime; it is here that the quadruple nature of the system has the greatest effect.
\end{enumerate}
In Figure \ref{fig:compare_regimes}, we compare the eccentricity excitation of $m_{1}$ between the modified LK and resonant regimes for trajectories with identical initial conditions. When $\beta \gg 1$, the maximal eccentricity $e_{1} \approx 0.95$ is consistent with Eq.~(\ref{eq:mod_LK_law}). When $\beta \sim 1$, however, extreme eccentricities $1 - e_{1} \sim 10^{-4}$ can be achieved even with moderate initial inclinations. We quantify this in the next section.

\subsection{Short-Range Forces and the `Extreme' LK Window} \label{s:Dyn:SRF}

For a planet initially at a large distance ($a_{1} \gtrsim 10 \AU$) to migrate to $\sim 0.02 \AU$ through the LK mechanism, extreme eccentricity excitation ($1 - e_{1} \lesssim 10^{-3}$) is required. However, when the planet's pericenter separation $a_{1} (1 - e_{1})$ from the host star is small, short-range forces (SRFs) can become significant. We implement these additional forces in our model following \citet{LML2015}.

We include SRFs arising from the 1PN correction to the gravitational potential of $m_{0}$ and from the tidal distortion of $m_{1}$. These contribute an additional term in Eq.~(\ref{eq:vecteqs_de1dt}):
\begin{equation} \label{eq:de1dt_SRFs}
    \left( \frac{\dif \vect{e}_{1}}{\dif t} \right)_{\rm SRF} = \left( \dot{\omega}_{\rm 1PN} + \dot{\omega}_{\rm tide} \right) \Lhat_{1} \times \vect{e}_{1},
\end{equation}
where
\begin{align}
    \dot{\omega}_{\rm 1PN} &= \frac{3 G m_{0}}{c^{2} a_{1}} \frac{n_{1}}{1-e_{1}^{2}}, \\
    \dot{\omega}_{\rm tide} &= \frac{15}{2} k_{2,1} \frac{m_{0}}{m_{1}} \left( \frac{R_{1}}{a_{1}} \right)^{5} \frac{1 + (3/2) e_{1}^{2} + (1/8) e_{1}^{4}}{(1 - e_{1}^{2})^{5}} n_{1},
\end{align}
with $R_{1}$ and $k_{2,1}$ the radius and tidal Love number of $m_{1}$.

\begin{figure*}
    \centering
    \includegraphics[width=0.95\textwidth]{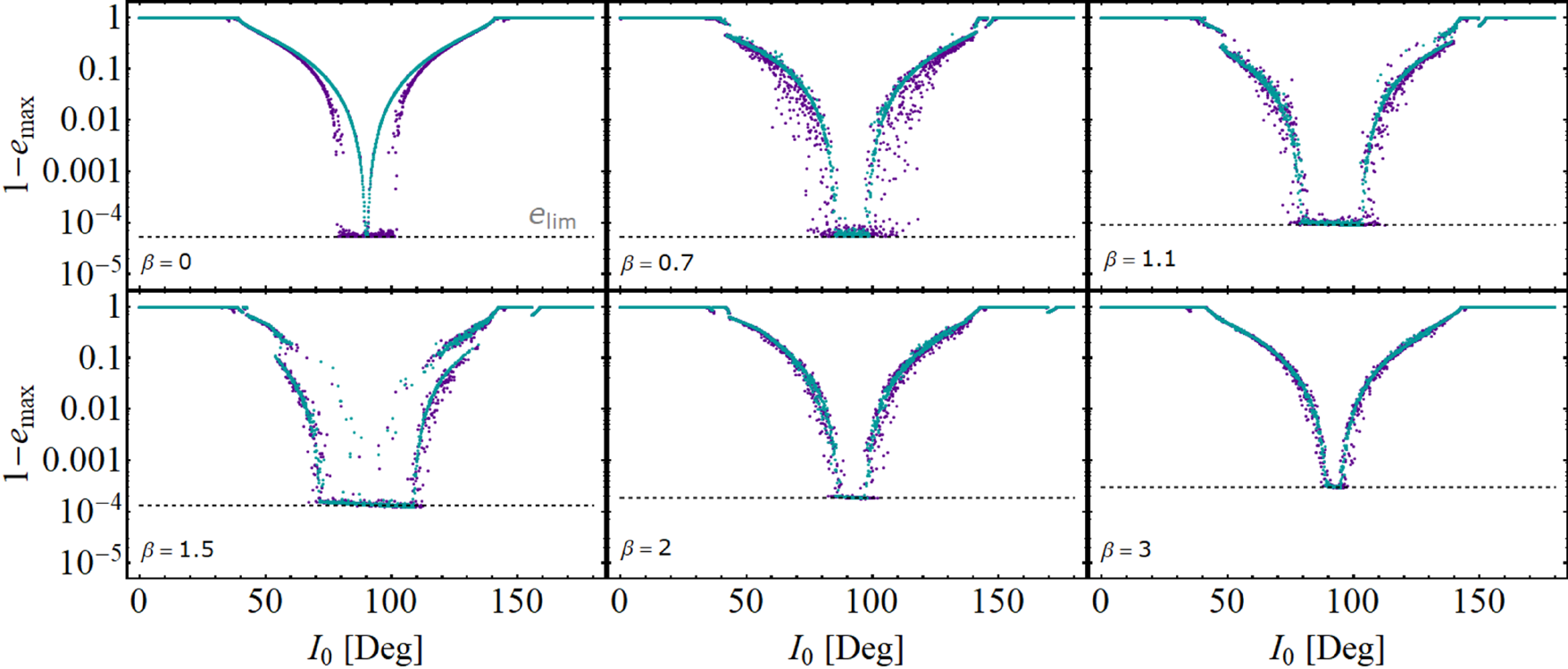}
    \caption{Maximum eccentricity of orbit 1 (as $1-e$) versus initial inclination $I_{0} \equiv I_{1,0}$ for different values of $\beta$ (Eq.~\ref{eq:beta_def}). In all cases, we choose $m_{0} = 0.6 \MSol$, $m_{1} = 1.0 \MJ$, $m_{2} = m_{3} = 0.33M_\odot$, $a_{2} = 60 \AU$ (except the top-left panel where $a_{2} = 1 \AU$), $e_{2} = 0$, $e_{3} = 0.3$, $a_{3} = 1500 \AU$, and $I_{2} = 30^{\circ}$ ($\alpha \approx 2^{\circ}$). The value of $a_{1}$ in each panel is $52 \AU$ ($\beta = 0$ and $\beta=0.7$), $38.5 \AU$ ($\beta=1.1$), $31.3 \AU$ ($\beta = 1.5$), $25.8 \AU$ ($\beta=2$), and $19.7 \AU$ ($\beta=3$). The planet's initial $e_{1,0} = 0.01$. The argument of pericentre $\omega_{1,0}$ and longitude of the node $\Omega_{1,0}$ are randomly chosen in $[0, 2\pi)$. We integrate the vectorial secular equations for $500 \tLK$ and plot the maximal eccentricity achieved by $m_{1}$ in that time. We include SRFs due to the 1PN correction and tidal distortion of $m_{1}$, assuming $R_{1} = 1.0 \RJ$ and $k_{2,1} = 0.37$. The cyan dots represent the results including quadrupole-order perturbations and SRFs whilst the purple dots also include the octupole-order terms from $m_{23}$ perturbing $m_{1}$. The dotted horizontal line is the limiting eccentricity $e_{\rm lim}$ of Eq.~(\ref{eq:elim_def}).}
    \label{fig:window}
\end{figure*}

SRFs impose an upper limit $e_{\rm lim}$ on the eccentricity that can be achieved through secular dynamical excitation; this limit is given by \citep{LML2015}
\begin{equation} \label{eq:elim_def}
    \left( \frac{\dot{\omega}_{\rm 1PN}}{\dot{\omega}_{\rm LK}} + \frac{1}{9} \frac{\dot{\omega}_{\rm tide}}{\dot{\omega}_{\rm LK}} \right)_{e_{1} = e_{\rm lim}} = \frac{9}{8}
\end{equation}
where
\begin{equation}
    \dot{\omega}_{\rm LK} = \frac{1}{\tLK (1 - e_{1}^{2})^{1/2}}.
\end{equation}
Eq.~(\ref{eq:elim_def}) was derived for the standard `2+1' quadrupole LK problem with SRFs, but it is also valid when octupole effects are included (\citealt{LML2015}; see also Section \ref{s:Dyn:Oct}). Our numerical calculations show that it holds for the `2+2' problem as well (Fig.~\ref{fig:window}).

In the general problem of HEM via the LK effect, additional SRFs can arise from the rotational distortion of $m_{0}$ and $m_{1}$. However, rotational effects are negligible compared to the 1PN and tidal effects for a giant planet migrating around a WD \citep[e.g.,][]{ASL2016}. The limiting eccentricity for our purposes is determined mainly by the tidal perturbation; thus:
\begin{align} \label{eq:elim_tidal}
    1-e_{\rm lim} &\simeq 5 \times 10^{-5} \left( \frac{m_{0}}{0.6 \MSol} \right)^{4/9} \left( \frac{m_{23}}{0.66 \MSol} \right)^{-2/9} \nonumber \\
    & \hspace{0.5cm} \times \left( \frac{k_{2,1}}{0.37} \right)^{2/9} \left( \frac{R_{1}}{\RJ} \right)^{10/9} \left( \frac{m_{1}}{\MJ} \right)^{-2/9} \nonumber \\
    & \hspace{0.5cm} \times \left( \frac{a_{3,{\rm eff}}}{1500 \AU} \right)^{2/3} \left( \frac{a_{1}}{50 \AU} \right)^{-16/9},
\end{align}
where we have assumed planetary properties analogous to Jupiter.

To produce the observed giant planet (at $a_{1,{\rm obs}} \approx 0.02 \AU$) around WD\,1856+534 via HEM, we require the pericentre distance $a_{1} (1 - e_{1})$ to reach below $0.01 \AU$ (recall that tidal dissipation conserves angular momentum). Using Eq.~(\ref{eq:elim_tidal}), we find that the semi-major axis must satisfy
\begin{align} \label{eq:a1min}
    a_{1} &\gtrsim 8.4 \AU \left( \frac{m_{0}}{0.6 \MSol} \right)^{4/7} \left( \frac{m_{23}}{0.66 \MSol} \right)^{-2/7} \nonumber \\
    & \hspace{0.5cm} \times \left( \frac{k_{2,1}}{0.37} \right)^{2/7} \left( \frac{R_{1}}{\RJ} \right)^{10/7} \left( \frac{m_{1}}{\MJ} \right)^{-2/7} \nonumber \\
    & \hspace{0.5cm} \times \left( \frac{a_{3,{\rm eff}}}{1500 \AU} \right)^{6/7} \left( \frac{a_{1,{\rm obs}}}{0.02 \AU} \right)^{-9/7},
\end{align}
in order for external perturbations to overcome SRFs and push the planet to sufficiently high eccentricity.

Even when Eq.~(\ref{eq:a1min}) is satisfied, the standard `2+1' LK effect can produce extreme eccentricity excitation only when $I_{1,0}$ is very close to 90 degrees (top-left panel of Fig.~\ref{fig:window}, cyan points). This is where the resonant LK effect in a 2+2 system becomes important: when $\beta \sim 1$, the inclination resonance broadens the LK window for extreme eccentricities \citep{HL2017}. For typical properties of the WD\,1856 system (from Table \ref{tab:properties}: $m_{0} = 0.6 \MSol$, $m_{2} = m_{3} = 0.33 \MSol$, $a_{2} = 60 \AU$, $a_{3} = 1500 \AU$, $I_{2} = 30^{\circ}$; note that $\alpha \simeq 2^{\circ}$ and $J \simeq L_{3}$), Eq.~(\ref{eq:beta_def}) provides the value of $a_{1}$ corresponding to a given $\beta$:
\begin{align} \label{eq:a1_beta}
    a_{1} &= \frac{49.5 \AU}{\beta^{2/3}} \left( \frac{m_{0}}{0.6 \MSol} \right) \left( \frac{m_{23}}{0.66 \MSol} \right)^{-1} \nonumber \\
    & \hspace{1.5cm} \times \left( \frac{a_{2}}{60 \AU} \right) \left( \frac{\cos I_{2}}{\sqrt{3}/2} \right)^{2/3}.
\end{align}

To evaluate the `extreme' LK window for a given $\beta$, we numerically integrate the quadrupole-order secular equations (including SRFs) for a duration of $500 \tLK$ using initial conditions over the full range of $I_{1,0}$ with the angles of the node ($\Omega_{1,0}$) and pericentre ($\omega_{1,0}$) randomly distributed. Fig.~\ref{fig:window} shows the result of this calculation as cyan points. For $\beta \ll 1$ (the quasi-LK regime), the second binary behaves like a point mass $m_{23}$, the relation between $I_{1,0}$ and $e_{1,{\rm max}}$ is analytical (Eq.~\ref{eq:std_LK_law}), and the limiting eccentricity is achieved only if the initial inclination is extremely close to $90^{\circ}$. For $\beta$ around unity, we see a substantial widening of the extreme LK window. The window appears to be widest around $\beta = 1.5$, spanning the range $70^{\circ} \lesssim I_{1,0} \lesssim 110^{\circ}$. For progressively larger $\beta$, the window shrinks again to a narrow range around $I_{1,0} = 90^{\circ}$ as the system enters the modified LK regime.

Importantly, Fig.~\ref{fig:window} confirms that despite the complex LK evolution for $\beta\sim 1$, the limiting eccentricity of Eqs.~(\ref{eq:elim_def}, \ref{eq:elim_tidal}) still sets the floor for $1-e_{1}$. Note that the ``double-valued'' features seen in Fig.~\ref{fig:window} (e.g., in the lower-left panel) are a consequence of chaotic evolution for $\beta \sim 1$: the time required to achieve large eccentricities (beyond the standard `2+1' LK effect) is highly variable (i.e. is sensitive to initial conditions) and may exceed $500 \tLK$ (the duration of each integration).

\subsection{Octupole-Order Effects} \label{s:Dyn:Oct}

So far, we have discussed the dynamics of a test particle in a 2+2 system at the quadrupole order of approximation. However, octupole-order effects may be necessary for an accurate treatment for nonzero $e_{3}$ and $a_{1}$ sufficiently large. The strength of the octupole perturbation of $m_{1}$ by $m_{23}$ relative to the quadrupole is measured by the dimensionless quantity
\begin{equation} \label{eq:epsoct_def}
    \varepsilon_{\rm oct} = \frac{a_{1}}{a_{3}} \frac{e_{3}}{1 - e_{3}^{2}}.
\end{equation}
\citet{LML2015} showed that including the octupole terms in the 2+1 LK problem with SRFs preserves the limiting eccentricity (Eq.~\ref{eq:elim_def}) and that octupole effects can enhance the extreme LK window. Orbit 1 can always achieve $e_{1} = e_{\rm lim}$ when the initial inclination $I_{1,0}$ exceeds a critical value $I_{1,{\rm cr}}$ (i.e., $I_{1,{\rm cr}} < I_{1,0} < 180^{\circ}-I_{1,{\rm cr}}$; see the top-left panel of Fig.~\ref{fig:window}). An analytical fit for $I_{1,{\rm cr}}$ is
\begin{align} \label{eq:Icrit_oct}
    \cos^{2} (I_{1,{\rm cr}}) &\simeq 0.26 \left( \frac{\varepsilon_{\rm oct}}{0.1} \right) - 0.536 \left( \frac{\varepsilon_{\rm oct}}{0.1} \right)^{2} \nonumber \\
    & \hspace{0.25cm} + 12.05 \left( \frac{\varepsilon_{\rm oct}}{0.1} \right)^{3} - 16.78 \left( \frac{\varepsilon_{\rm oct}}{0.1} \right)^{4}
\end{align}
for $\varepsilon_{\rm oct} \lesssim 0.05$ and $\cos^{2} (I_{1,{\rm cr}}) \simeq 0.45$ for $\varepsilon_{\rm oct} \gtrsim 0.05$ \citep{MLL2016}. This form of enhancement is independent of the secular resonance effect in a 2+2 system and therefore can persist when $\beta \ll 1$. For fiducial properties of the WD\,1856 system ($a_{3} = 1500 \AU, e_{3} = 0.3$), we have
\begin{align}
    \varepsilon_{\rm oct} &\simeq 0.01 \left( \frac{30 a_{1}}{a_{3}} \right) \left( \frac{e_{3}}{0.3 (1-e_{3}^{2})} \right), \label{eq:epsoct_numbers} \\
    |I_{1,{\rm cr}} - 90^{\circ}| &\simeq 10^\circ \left( \frac{30 a_{1}}{a_{3}} \right)^{1/2} \left( \frac{e_3}{0.3 (1-e_{3}^{2})} \right)^{1/2}. \label{eq:I1cr_numbers}
\end{align}

In order to evaluate the relative contributions of the octupole and 2+2 ``resonant'' effects to the enhancement of LK window, we repeat the exercise of Section \ref{s:Dyn:SRF}, this time including additional terms in Eqs.~(\ref{eq:vecteqs_dj1dt})--(\ref{eq:vecteqs_dL3dt}) to describe the octupole-order perturbation of orbit 1 by orbit 3 following \citet{LML2015} \citep[see also][]{LL2019}. The results are displayed as purple points in Fig.~\ref{fig:window}. As predicted, the octupole-order results feature an extreme LK window even for $\beta \ll 1$, consistent with Eq.~(\ref{eq:I1cr_numbers}).

For $\beta = 0.7$, the inclusion of octupole effects significantly widens the LK window relative to 2+2 resonant quadrupole effects alone. This suggests that the resonant-quadrupole and octupole effects can interfere constructively for $0.1 \lesssim \beta \lesssim 1$. For $\beta \gtrsim 1$, $a_{1}$ and $\varepsilon_{\rm oct}$ are smaller and the ``resonant'' quadrupole effect dominates the width of the extreme LK window; in these cases, octupole effects introduce a modest scatter in $e_{1,{\rm max}}$ about the quadrupole result but do not otherwise affect the outcome.

\section{High-Eccentricity Migration around WD~1856+534} \label{s:HEM_ex}

We now apply our model to the WD\,1856+534 system in order to demonstrate the feasibility of HEM through the enhanced LK effect in the resonant regime. For simplicity, we do not simulate the system's dynamical evolution prior to the WD phase (but see Section \ref{s:Disc:preWD}). We also neglect octupole effects in this section.

We assume the same stellar and planetary parameters as in the previous section. In addition to the quadrupole secular perturbations and SRFs used previously, we now include dissipation of the equilibrium tide raised on the planet by the WD. We adopt the weak-friction model, where the planet's internal dissipation is parametrized by a constant lag-time $\Delta t_{\rm L,1}$ \citep[e.g.,][]{Alexander1973,Hut1981}. The additional terms in Eqs.~(\ref{eq:vecteqs_dj1dt}, \ref{eq:vecteqs_de1dt}) due to weak friction are given by \citet{ASL2016}. We assume that the planet rotates pseudosynchronously during HEM.

\begin{figure*}
    \centering
    \includegraphics[width=0.9\textwidth]{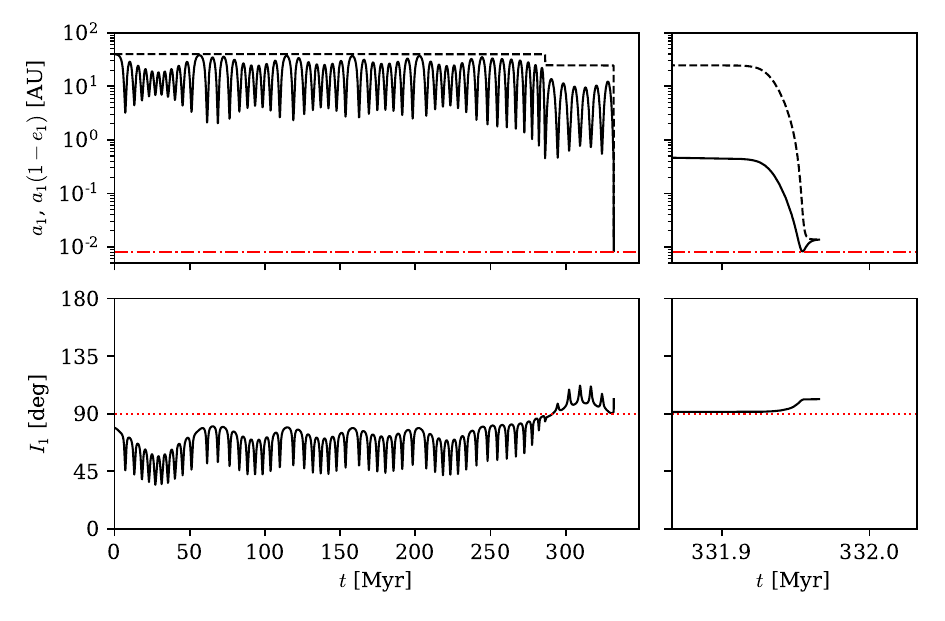}
    \caption{An example of HEM of a Jupiter-like planet in the WD\,1856 system through the resonant `2+2' LK effect. The upper panels show the semi-major axis ($a_{1}$, dashed black curve) and pericentre distance ($a_{1} [1-e_{1}]$, solid black) with the tidal disruption radius $\approx 2 R_{1} (m_{0}/m_{1})^{1/3}$ (horizontal red line) marked for reference. The lower panels show the inclination $I_{1}$ (black curve) with $I_{1} = 90^{\circ}$ (in red) for reference. The left panels show the full integration over $\sim 332 \Myr$ whilst the right panels zoom in on the final circularization phase. The parameters are: WD mass $m_{0} = 0.6 \MSol$, planetary mass and radius $m_{1} = 1.0 \MJ$ and $R_{1} = 1.0 \RJ$, companion masses $m_{2} = m_{3} = 0.33 \MSol$, $a_{2} = 60 \AU$, $a_{3} = 1500 \AU$, $e_{3} = 0.3$, and $I_{2} = 30^{\circ}$, corresponding to $\alpha \approx 2^{\circ}$ and $\beta = 1.11$. The planet's initial conditions are $a_{1} = 40 \AU$, $e_{1} = 0.01$, $I_{1} = 75^{\circ}$, $\Omega_{1} = 289 \fdg 7$, and $\omega_{1} = 175 \fdg 4$, and the tidal Love number and lag-time are $k_{2,1} = 0.37$ and $\Delta t_{\rm L,1} = 10 \, {\rm s}$. Note that the planet's final orbit is retrograde and nearly perpendicular to orbit 3.}
    \label{fig:WD1856_example}
\end{figure*}

In Figure \ref{fig:WD1856_example}, we display an example of successful HEM of WD\,1856\,b from an initial configuration with $a_{1} = 40 \AU$ (corresponding to $\beta = 1.11$) and $I_{1} = 75^{\circ}$. For this example, we adopt $\Delta t_{\rm L,1} = 10 \, {\rm s}$, corresponding to a factor of 100 enhancement relative to Jupiter's dissipation; this is necessary to ensure timely orbital circularization of the planet and avoid tidal disruption (see Section \ref{s:Disc:Cav}). The planet's minimal pericentre distance is just larger than the tidal disruption limit,
\begin{equation} \label{eq:rpdis}
    r_{\rm p,dis} = \eta R_{1} \left( \frac{m_{0}}{m_{1}} \right)^{1/3}
\end{equation}
where $\eta \approx 2$--$3$ for a giant planet \citep[e.g.,][]{GRL2011}. The planet's final semi-major axis is approximately twice this value, close to WD\,1856\,b's orbit at $\approx 0.02 \AU$. The entire evolution takes place in slightly less than $332 \Myr$, well within the $\sim 6 \Gyr$ cooling age of WD\,1856 \citep{Vanderburg+2020}.

We note in the bottom panels of Fig.~\ref{fig:WD1856_example} that the planet's final orbit is nearly perpendicular to the outer orbit. This is consistent with the actual viewing geometry for this system, with orbit 3 lying roughly in the celestial plane and orbit 1 along the line of sight \citep{Vanderburg+2020}.

\section{Pre-WD Dynamical Evolution} \label{s:Disc:preWD}

Let us consider the possible dynamical evolution of the WD\,1856 system prior to the WD phase. As we noted previously, the transiting planet could not have migrated to its current location until after the host star had evolved into a WD; if an enhanced LK effect was responsible for this migration, then it is desirable that the effect be suppressed prior to the WD phase. The two most likely ways to accomplish this involve stellar evolution or the presence of additional planets in this system prior to the WD phase.

Most known WDs that show evidence for planetary systems evolved from main-sequence (MS) stars somewhat more massive than the Sun, typically $\sim 1.5$--$3.0 \MSol$ \citep{KGF2014}. Such stars become WDs of $\sim 0.5$--$0.8 \MSol$ \citep{Kalirai+2008,Choi+2016}, losing most of their mass during the AGB phase. AGB mass loss occurs over several Myr, a long time-scale compared to the orbital period of planetary or stellar companions closer than $\sim 10^{4} \AU$; thus its effect on a 2+2 system can be treated using the principle of adiabatic invariance.

Consider the adiabatic reduction of the primary mass from $m_{0,{\rm MS}}$ (the MS value) to $m_{0} = f m_{0,{\rm MS}}$ (the final WD mass, with $f < 1$). Assuming that mass is expelled isotropically and that none is captured by the other bodies, adiabatic invariance implies that orbits 1 and 3 expand according to
\begin{align}
    a_{1,{\rm MS}} &\to a_{1} = \frac{a_{1,{\rm MS}}}{f}, \\
    a_{3,{\rm MS}} &\to a_{3} =  \frac{m_{0,{\rm MS}} + m_{23}}{m_{0} + m_{23}} a_{3,{\rm MS}}
\end{align}
whilst all other orbital elements are unchanged. It follows that the LK time-scale and the parameter $\beta$ change according to
\begin{align}
    t_{\rm LK,MS} &\to \tLK = f^{2} \left( \frac{m_{0,{\rm MS}} + m_{23}}{m_{0} + m_{23}} \right)^{3} t_{\rm LK,MS}, \label{eq:tLK_adiab} \\
    \beta_{\rm MS} &\to \beta = f^{3} \beta_{\rm MS}, \label{eq:beta_adiab}
\end{align}
where we have used $J \simeq L_{3}$ and $I_{2} = {\rm constant}$ during the evolution. We see that adiabatic mass loss reduces $\beta$ by a large factor. Thus, orbit 1 can transition from the modified LK regime ($\beta_{\rm MS} \gg 1$) to the resonant regime ($\beta \sim 1$) as a result of stellar mass loss.

For typical WD\,1856 system parameters, significant enhancement of the extreme LK window requires $2 \gtrsim \beta \gtrsim 0.7$ (Fig.~\ref{fig:window}),  corresponding to the planet's semi-major axis at the end of the AGB phase in the range $28 \AU \lesssim a_1 \lesssim 57 \AU$ (Eq.~\ref{eq:a1_beta}). If we adopt $m_{0} = 0.6 \MSol$ for the WD and $m_{0,\rm {MS}} = 2.0 \MSol$ for the progenitor \citep[per the initial--final mass relation of][]{Kalirai+2008}, i.e. $f = 0.3$, we find that this ``resonant'' $a_{1}$ range translates to $8 \AU \lesssim a_{1,\rm {MS}} \lesssim 17 \AU$. This is consistent with WD\,1856\,b having been a typical ``cold Jupiter'' during the MS.

As discussed in Section \ref{s:Dyn:Oct}, octupole-order effects can also enhance the window for extreme eccentricity excitation through the LK effect. During adiabatic mass loss, $\varepsilon_{\rm oct}$ (Eq.~\ref{eq:epsoct_def}) evolves according to
\begin{equation}
    \varepsilon_{\rm oct,MS} \to \varepsilon_{\rm oct} = \frac{1}{f} \left( \frac{m_{0} + m_{23}}{m_{0,{\rm MS}} + m_{23}} \right) \varepsilon_{\rm oct,MS}.
\end{equation}
For the canonical masses in the WD\,1856 system and $f = 0.3$, we have $\varepsilon_{\rm oct,MS} \approx 0.63 \varepsilon_{\rm oct}$, meaning octupole effects are less important during the MS \citep*[see also][]{SNG2020}. From Eqs.~(\ref{eq:epsoct_numbers}, \ref{eq:I1cr_numbers}), octupole effects can induce an appreciable extreme LK window (e.g., wider than $10^{\circ}$) only if the planet's semi-major axis is larger than $\sim 1/30$ of the semi-major axis of orbit 3, or $a_{1,{\rm MS}} \gtrsim 24 \AU$ for $a_{3,\rm {MS}} = 700 \AU$.

In principle, HEM through the LK effect could occur during the MS. Equation (\ref{eq:elim_tidal}) with the above MS system parameters implies that a planet at $a_{1,{\rm MS}} \approx 15 \AU$ could have reached a pericentre distance $a_{1,{\rm MS}} (1 - e_{\rm lim}) \approx 0.007 \AU$. However, the inclination window to reach such extreme eccentricities would have been narrow since $\beta_{\rm MS} \gg 1$ and $\varepsilon_{\rm oct,MS} \ll 1$. AGB mass loss expands the extreme LK window, acting as a ``trigger'' for HEM during the WD phase.

In addition to the effects of stellar evolution, another planet orbiting closer to the host star could have suppressed the LK effect during the MS by augmenting the free precession rate of WD\,1856\,b \citep[see][]{HTT1997,PM2017}. A planet of mass $m'$ at a distance $a' < a_{1,{\rm MS}}$ from the host star would completely suppress LK oscillations of $m_{1}$ if
\begin{align}
    m' &\gtrsim m_{23} \frac{a_{1,{\rm MS}}^{5}}{a_{3,{\rm MS}}^{3} (a')^{2}} \nonumber \\
    &\approx 0.2 \MJ \left( \frac{a_{1,{\rm MS}}}{10 \AU} \right)^{5} \left( \frac{a_{3,{\rm MS}}}{700 \AU} \right)^{-3} \left( \frac{a'}{1 \AU} \right)^{-2}.
\end{align}
A smaller $m'$ would still be sufficient to suppress high-eccentricity excursions of $m_{1}$. If this planet had $a' \lesssim 3 \AU$ during the MS, it would have been engulfed during the AGB stage \citep[e.g.,][]{MV2012}, thereby ``switching on'' the LK effect during the WD phase. The architecture envisioned in this scenario (i.e., two planets between $\sim 1$ and $20 \AU$) is plausible given current knowledge of the population of extrasolar planets. For instance, \citet{Bryan+2016} found that $\sim 50 \%$ of giant planets orbiting between $1$ and $5 \AU$ have companions of comparable mass within a distance of $20 \AU$.

\section{Main Caveats and Uncertainties} \label{s:Disc:Cav}

\subsection{Tidal Disruption and Survival of Migrating Planets} \label{s:Disc:Cav:rates}

The challenges of HEM through secular dynamics have been studied in the context of hot-Jupiter formation around MS stars. A major issue is that it is easy for a migrating planet to be tidally disrupted. Population-synthesis models \citep{Petrovich2015,ASL2016,Hamers+2017,VLA2019,TLV2019} and analytical calculations \citep{MLL2016} show that tidal disruption can limit the efficiency of hot-Jupiter formation through secular HEM to a few percent. In particular, although octupole effects broaden the `extreme' LK window and thus increase the migration fraction, most migrating planets are tidally disrupted \citep{ASL2016,MLL2016}. A similar situation occurs for the secular-chaos scenario \citep{WL2011,TLV2019}. In Section \ref{s:HEM_ex}, although we have not systematically surveyed a large parameter space, we also find that a large faction of migrating planets are disrupted for HEM in 2+2 systems. In fact, we had to use a relatively large tidal lag-time ($\Delta t_{\rm L,1} = 10 \, {\rm s}$) in order to find a reasonable number of surviving cases.

The problem of tidal disruption during HEM can be somewhat alleviated by invoking forms of tidal dissipation other than weak friction, such as chaotic dynamical tides with non-linear dissipation \citep{VL2018,Wu2018,VLA2019,TLV2019,VF2019}. \citet{VLA2019} show that the strong dissipation associated with chaotic tides can shepherd to safety some planets that are otherwise destined for tidal disruption by rapidly decreasing their eccentricities; they find $\sim 20\%$ of migrating planets survive with short final orbital periods.

Strong planet--planet scattering may also give rise to HEM. A planetary system that is dynamically stable during the MS can become unstable after AGB stellar mass loss \citep{DS2002}. However, strong scattering of unstable giant planets mostly results in ejections or planetary mergers. The `branching ratio' of one planet being injected into a low-pericentre orbit suitable for HEM is small \citep[e.g.,][]{MVV2014,VG2015,Veras+2016,Anderson+2020,Li+2020} unless the initial number of unstable planets is large \citep{Maldonado+2020}. We therefore consider this a less promising HEM mechanism.

\subsection{Properties of WD and Planet} \label{s:Disc:Cav:props}

We have assumed fiducial masses of $m_{0} = 0.6 \MSol$ and $m_{1} = 1.0 \MJ$ for the WD and planet, respectively. \citet{Vanderburg+2020} report a somewhat smaller WD mass of $(0.52 \pm 0.06) \MSol$ and obtain only an upper bound of $13.8 \MJ$ for the planet. The exact masses one assumes do not affect the key conclusions of this work as far as the LK effect is concerned. Since WD\,1856\,b has at most $\sim 2\%$ the mass of its host, the test-particle approximation is reasonable. Decreasing $m_{0}$ or increasing $m_{1}$ weakly affects the limiting eccentricity (Eq.~\ref{eq:elim_tidal}), shifting the minimal $a_{1}$ for migration (Eq.~\ref{eq:a1min}), and the tidal disruption distance (Eq.~\ref{eq:rpdis}) slightly inward. The range of $a_{1}$ for which the 2+2 resonance is active also shifts slightly inward for decreased $m_{0}$ (Eq.~\ref{eq:a1_beta}). The tidal circularization timescale, which is proportional to $m_{1} / (m_0)^{2}$ in the weak-friction theory, is affected more significantly. If the planet's true mass is $10 \MJ$, then stronger dissipation ($\Delta t_{\rm L,1} \gtrsim 10^{2} \, {\rm s}$) would be required for the planet's migration and survival (but see Section \ref{s:Disc:Cav:rates} for a discussion of chaotic dynamical tides).

The values of the mass and cooling age of WD\,1856 reported by \citet{Vanderburg+2020} were obtained by fitting atmospheric models to the stellar spectrum. Taken at face value, the best-fitting model -- with a WD mass of $0.52 \MSol$ and a cooling age of $\sim 6 \Gyr$ -- implies a roughly solar-mass progenitor that spent $\sim 10 \Gyr$ on the MS. However, this is at odds with the system's likely membership in the Galactic thin disc, which would imply that the system is younger than $\sim 10 \Gyr$ in total. \citet{Lagos+2020} have suggested a resolution to this conundrum in which WD\,1856\,b migrated through common-envelope evolution, reducing the final mass of the WD relative to its progenitor in the process \citep[see also][]{LS1984}. On the other hand, a modest systematic error in the estimated WD mass, perhaps due to the difficulty of fitting WD\,1856's featureless spectrum, can also eliminate the age discrepancy without invoking common-envelope evolution \citep{Vanderburg+2020}. Our assumption about the WD and progenitor masses aligns with the latter explanation.

\subsection{Occurrence Rates} \label{s:Disc:Cav:occur}

The observed configuration of WD\,1856\,b is improbable: the planet only obscures half of the WD in transit despite having $\sim 10$ times its radius \citep{Vanderburg+2020}. The probability of observing such `grazing' transits is $\sim R_{0}/a_{1,{\rm obs}} = 0.2\%$ (for $a_{1,{\rm obs}} = 0.02 \AU$ and WD radius $R_{0} \approx \RE$) whilst the probability of a `full' transit is $\sim \RJ / a_{\rm 1,obs} = 2\%$. Hence, the detection of WD\,1856\,b with a `grazing' transit suggests there could be $\sim 10$ times as many systems where the WD is completely blocked. However, no other transit of a WD by a giant planet has been detected by TESS thus far, suggesting that the transit geometry of WD\,1856\,b is ``lucky'' (A.~Vanderburg, private communication). Given that TESS has observed $\sim 1700$ isolated WDs thus far \citep{Vanderburg+2020}, this would imply that short-period giant planets could orbit $\sim 2$--$3 \%$ of them. This agrees with the $2 \sigma$ upper limit obtained from the WD sample observed by {\it Kepler}/K2 \citep{vSvE2018} but is somewhat higher than that from Pan-STARRS \citep{Fulton+2014}. The discovery of additional planets transiting WDs will eventually allow an assessment of whether this population can be attributed to HEM alone.

\section{Conclusion} \label{s:Sum}

In this paper, we have demonstrated the possible HEM of WD\,1856\,b through an enhanced LK effect from the WD's distant M-dwarf companions. We show that the inclination resonance in hierarchical 2+2 systems and the octupole effect can both significantly broaden the LK window for extreme eccentricity excitation (see Fig.~\ref{fig:window}), allowing HEM to operate in the WD\,1856 system for a wide range of initial planetary orbital inclinations. The requirement that secular perturbations from the companion stars be able to overcome SRFs imposes an absolute limit of $a_{1} \gtrsim 8 \AU$ for the planet's semi-major axis just prior to migration. Due to the large enhancement of the extreme LK window by the 2+2 resonance, we find that the planet most likely migrated from $30 \AU \lesssim a_{1} \lesssim 60 \AU$, corresponding to $\sim 10$--$20 \AU$ during the host's MS phase. Importantly, extreme eccentricity excitation through the 2+2 resonance is delayed until the WD phase at this distance.

Although WD\,1856\,b is so far the only intact short-period giant planet known to orbit a WD, HEM and tidal disruption of giant planets may have occurred in other WD systems. For instance, WD\,J0914+1914 possesses a gaseous accretion disc and a pollution signature rich in hydrogen, oxygen, and sulfur \citep{Gansicke+2019}. This has been interpreted as evidence for an ice giant's ongoing or previous disruption by the WD. As noted in Section \ref{s:Disc:Cav}, any flavor of HEM will more likely lead to tidal disruption than survival of the planet. Although WD\,J0914+1914's cooling age is quite short ($\sim 13 \Myr$), secular interactions can deliver a planet to the WD in time given a suitable companion \citep{SNG2020}.

Atmospheric pollution may be a more prevalent signature of remnant planetary systems around WDs than the occurrence of short-period giant planets. To advance understanding of the occurrence of planetary systems in multiple-star systems, we recommend further observational efforts to determine the fraction of polluted WDs that belong to wide binaries or hierarchical triples. Previous studies by \citet{Zuckerman2014} and \citet{Wilson+2019} find consistent pollution fractions among WDs occurring singly and in wide binaries. The {\em Gaia} mission and LSST will aid in the identification of previously unknown stellar or substellar companions of WDs \citep[e.g.,][]{GF+2019}, offering the opportunity to refine these results.

After the original submission of this work, other studies of LK migration of WD\,1856\,b were independently put forth \citep{MP2020,SNG2020}. These works did not consider the full `2+2' LK effect but rather focused on the `2+1' octupole effect (see Section \ref{s:Dyn:Oct}). \citet{MP2020} argue for an initial semi-major axis $a_{1,{\rm MS}} \sim 2$--$3 \AU$ (or $a_{1} \sim 6$--$10 \AU$ just before migration), as a planet with greater $a_{1}$ would experience a stronger octupole LK effect and therefore might be tidally disrupted (assuming the weak-friction tidal theory). However, migration from $a_{1} \lesssim 10 \AU$ can occur only when the initial inclination is close to $90^{\circ}$ (Eq.~\ref{eq:Icrit_oct}), which is relatively unlikely. The problem of tidal disruption may be alleviated by chaotic tides (see Section \ref{s:Disc:Cav:rates}). Meanwhile, \citet{SNG2020} favour an initial orbit $20 \AU \lesssim a_{1,{\rm MS}} \lesssim 100 \AU$, for which the 2+1 octupole effect can excite extreme eccentricities for a wide range of inclinations. Since they did not account for the resonant effect in 2+2 systems, one would not expect planets to migrate from $a_{1,{\rm MS}} \lesssim 20 \AU$ in their calculations. Planets on such large orbits may also experience a strong octupole effect during the MS, depending on their initial inclination; the range of inclinations in which octupole-driven HEM is {\it forbidden} during the host's MS phase but {\it possible} during the WD phase is somewhat narrow. Finally, we note that there is some evidence that the giant-planet occurrence rate declines for $a_{1,{\rm MS}} \gtrsim 20 \AU$ \citep[e.g,][]{Bryan+2016,Vigan+2017}.

\section*{Acknowledgements}

CEO thanks Alexander Stephan, Dimitri Veras, and Laetitia Rodet for helpful discussions and the anonymous referee for comments on the manuscript. This work has been supported in part by National Science Foundation grant AST-1715246. This research has made use of NASA's Astrophysics Data System and of the software packages {\sc matplotlib} \citep{Hunter2007}, {\sc numpy} \citep{vDW+2011}, and {\sc scipy} \citep{Virtanen+2020}.

\section*{Data availability}

The data underlying this article will be shared on reasonable request to the corresponding author.

\bibliography{mn_20_4124_final.bib}

\begin{thebibliography}{}
\makeatletter
\relax
\def\mn@urlcharsother{\let\do\@makeother \do\$\do\&\do\#\do\^\do\_\do\%\do\~}
\def\mn@doi{\begingroup\mn@urlcharsother \@ifnextchar [ {\mn@doi@}
  {\mn@doi@[]}}
\def\mn@doi@[#1]#2{\def\@tempa{#1}\ifx\@tempa\@empty \href
  {http://dx.doi.org/#2} {doi:#2}\else \href {http://dx.doi.org/#2} {#1}\fi
  \endgroup}
\def\mn@eprint#1#2{\mn@eprint@#1:#2::\@nil}
\def\mn@eprint@arXiv#1{\href {http://arxiv.org/abs/#1} {{\tt arXiv:#1}}}
\def\mn@eprint@dblp#1{\href {http://dblp.uni-trier.de/rec/bibtex/#1.xml}
  {dblp:#1}}
\def\mn@eprint@#1:#2:#3:#4\@nil{\def\@tempa {#1}\def\@tempb {#2}\def\@tempc
  {#3}\ifx \@tempc \@empty \let \@tempc \@tempb \let \@tempb \@tempa \fi \ifx
  \@tempb \@empty \def\@tempb {arXiv}\fi \@ifundefined
  {mn@eprint@\@tempb}{\@tempb:\@tempc}{\expandafter \expandafter \csname
  mn@eprint@\@tempb\endcsname \expandafter{\@tempc}}}

\bibitem[\protect\citeauthoryear{{Alexander}}{{Alexander}}{1973}]{Alexander1973}
{Alexander} M.~E.,  1973, \mn@doi [\apss] {10.1007/BF00645172}, \href
  {https://ui.adsabs.harvard.edu/abs/1973Ap&SS..23..459A} {23, 459}

\bibitem[\protect\citeauthoryear{{Anderson}, {Storch}  \& {Lai}}{{Anderson}
  et~al.}{2016}]{ASL2016}
{Anderson} K.~R.,  {Storch} N.~I.,   {Lai} D.,  2016, \mn@doi [\mnras]
  {10.1093/mnras/stv2906}, \href
  {https://ui.adsabs.harvard.edu/abs/2016MNRAS.456.3671A} {456, 3671}

\bibitem[\protect\citeauthoryear{{Anderson}, {Lai}  \& {Pu}}{{Anderson}
  et~al.}{2020}]{Anderson+2020}
{Anderson} K.~R.,  {Lai} D.,   {Pu} B.,  2020, \mn@doi [\mnras]
  {10.1093/mnras/stz3119}, \href
  {https://ui.adsabs.harvard.edu/abs/2020MNRAS.491.1369A} {491, 1369}

\bibitem[\protect\citeauthoryear{{Bonsor} \& {Veras}}{{Bonsor} \&
  {Veras}}{2015}]{BV2015}
{Bonsor} A.,  {Veras} D.,  2015, \mn@doi [\mnras] {10.1093/mnras/stv1913},
  \href {https://ui.adsabs.harvard.edu/abs/2015MNRAS.454...53B} {454, 53}

\bibitem[\protect\citeauthoryear{{Bryan} et~al.,}{{Bryan}
  et~al.}{2016}]{Bryan+2016}
{Bryan} M.~L.,  et~al., 2016, \mn@doi [\apj] {10.3847/0004-637X/821/2/89},
  \href {https://ui.adsabs.harvard.edu/abs/2016ApJ...821...89B} {821, 89}

\bibitem[\protect\citeauthoryear{{Choi}, {Dotter}, {Conroy}, {Cantiello},
  {Paxton}  \& {Johnson}}{{Choi} et~al.}{2016}]{Choi+2016}
{Choi} J.,  {Dotter} A.,  {Conroy} C.,  {Cantiello} M.,  {Paxton} B.,
  {Johnson} B.~D.,  2016, \mn@doi [\apj] {10.3847/0004-637X/823/2/102}, \href
  {https://ui.adsabs.harvard.edu/abs/2016ApJ...823..102C} {823, 102}

\bibitem[\protect\citeauthoryear{{Dawson} \& {Johnson}}{{Dawson} \&
  {Johnson}}{2018}]{DJ2018}
{Dawson} R.~I.,  {Johnson} J.~A.,  2018, \mn@doi [\araa]
  {10.1146/annurev-astro-081817-051853}, \href
  {https://ui.adsabs.harvard.edu/abs/2018ARA&A..56..175D} {56, 175}

\bibitem[\protect\citeauthoryear{{Debes} \& {Sigurdsson}}{{Debes} \&
  {Sigurdsson}}{2002}]{DS2002}
{Debes} J.~H.,  {Sigurdsson} S.,  2002, \mn@doi [\apj] {10.1086/340291}, \href
  {https://ui.adsabs.harvard.edu/abs/2002ApJ...572..556D} {572, 556}

\bibitem[\protect\citeauthoryear{{Debes}, {Walsh}  \& {Stark}}{{Debes}
  et~al.}{2012}]{DWS2012}
{Debes} J.~H.,  {Walsh} K.~J.,   {Stark} C.,  2012, \mn@doi [\apj]
  {10.1088/0004-637X/747/2/148}, \href
  {https://ui.adsabs.harvard.edu/abs/2012ApJ...747..148D} {747, 148}

\bibitem[\protect\citeauthoryear{{Fabrycky} \& {Tremaine}}{{Fabrycky} \&
  {Tremaine}}{2007}]{FT2007}
{Fabrycky} D.,  {Tremaine} S.,  2007, \mn@doi [\apj] {10.1086/521702}, \href
  {https://ui.adsabs.harvard.edu/abs/2007ApJ...669.1298F} {669, 1298}

\bibitem[\protect\citeauthoryear{{Fang}, {Thompson}  \& {Hirata}}{{Fang}
  et~al.}{2018}]{FTH2018}
{Fang} X.,  {Thompson} T.~A.,   {Hirata} C.~M.,  2018, \mn@doi [\mnras]
  {10.1093/mnras/sty472}, \href
  {https://ui.adsabs.harvard.edu/abs/2018MNRAS.476.4234F} {476, 4234}

\bibitem[\protect\citeauthoryear{{Farihi}, {Jura}  \& {Zuckerman}}{{Farihi}
  et~al.}{2009}]{FJZ2009}
{Farihi} J.,  {Jura} M.,   {Zuckerman} B.,  2009, \mn@doi [\apj]
  {10.1088/0004-637X/694/2/805}, \href
  {https://ui.adsabs.harvard.edu/abs/2009ApJ...694..805F} {694, 805}

\bibitem[\protect\citeauthoryear{{Frewen} \& {Hansen}}{{Frewen} \&
  {Hansen}}{2014}]{FH2014}
{Frewen} S.~F.~N.,  {Hansen} B.~M.~S.,  2014, \mn@doi [\mnras]
  {10.1093/mnras/stu097}, \href
  {https://ui.adsabs.harvard.edu/abs/2014MNRAS.439.2442F} {439, 2442}

\bibitem[\protect\citeauthoryear{{Fulton} et~al.,}{{Fulton}
  et~al.}{2014}]{Fulton+2014}
{Fulton} B.~J.,  et~al., 2014, \mn@doi [\apj] {10.1088/0004-637X/796/2/114},
  \href {https://ui.adsabs.harvard.edu/abs/2014ApJ...796..114F} {796, 114}

\bibitem[\protect\citeauthoryear{{G{\"a}nsicke}, {Schreiber}, {Toloza},
  {Fusillo}, {Koester}  \& {Manser}}{{G{\"a}nsicke}
  et~al.}{2019}]{Gansicke+2019}
{G{\"a}nsicke} B.~T.,  {Schreiber} M.~R.,  {Toloza} O.,  {Fusillo} N. P.~G.,
  {Koester} D.,   {Manser} C.~J.,  2019, \mn@doi [\nat]
  {10.1038/s41586-019-1789-8}, \href
  {https://ui.adsabs.harvard.edu/abs/2019Natur.576...61G} {576, 61}

\bibitem[\protect\citeauthoryear{{Gentile Fusillo} et~al.,}{{Gentile Fusillo}
  et~al.}{2019}]{GF+2019}
{Gentile Fusillo} N.~P.,  et~al., 2019, \mn@doi [\mnras]
  {10.1093/mnras/sty3016}, \href
  {https://ui.adsabs.harvard.edu/abs/2019MNRAS.482.4570G} {482, 4570}

\bibitem[\protect\citeauthoryear{{Guillochon}, {Ramirez-Ruiz}  \&
  {Lin}}{{Guillochon} et~al.}{2011}]{GRL2011}
{Guillochon} J.,  {Ramirez-Ruiz} E.,   {Lin} D.,  2011, \mn@doi [\apj]
  {10.1088/0004-637X/732/2/74}, \href
  {https://ui.adsabs.harvard.edu/abs/2011ApJ...732...74G} {732, 74}

\bibitem[\protect\citeauthoryear{{Hamers} \& {Lai}}{{Hamers} \&
  {Lai}}{2017}]{HL2017}
{Hamers} A.~S.,  {Lai} D.,  2017, \mn@doi [\mnras] {10.1093/mnras/stx1319},
  \href {https://ui.adsabs.harvard.edu/abs/2017MNRAS.470.1657H} {470, 1657}

\bibitem[\protect\citeauthoryear{{Hamers} \& {Portegies Zwart}}{{Hamers} \&
  {Portegies Zwart}}{2016a}]{HPZ2016a}
{Hamers} A.~S.,  {Portegies Zwart} S.~F.,  2016a, \mn@doi [\mnras]
  {10.1093/mnras/stw784}, \href
  {https://ui.adsabs.harvard.edu/abs/2016MNRAS.459.2827H} {459, 2827}

\bibitem[\protect\citeauthoryear{{Hamers} \& {Portegies Zwart}}{{Hamers} \&
  {Portegies Zwart}}{2016b}]{HPZ2016b}
{Hamers} A.~S.,  {Portegies Zwart} S.~F.,  2016b, \mn@doi [\mnras]
  {10.1093/mnrasl/slw134}, \href
  {https://ui.adsabs.harvard.edu/abs/2016MNRAS.462L..84H} {462, L84}

\bibitem[\protect\citeauthoryear{{Hamers}, {Perets}, {Antonini}  \& {Portegies
  Zwart}}{{Hamers} et~al.}{2015}]{Hamers+2015}
{Hamers} A.~S.,  {Perets} H.~B.,  {Antonini} F.,   {Portegies Zwart} S.~F.,
  2015, \mn@doi [\mnras] {10.1093/mnras/stv452}, \href
  {https://ui.adsabs.harvard.edu/abs/2015MNRAS.449.4221H} {449, 4221}

\bibitem[\protect\citeauthoryear{{Hamers}, {Antonini}, {Lithwick}, {Perets}  \&
  {Portegies Zwart}}{{Hamers} et~al.}{2017}]{Hamers+2017}
{Hamers} A.~S.,  {Antonini} F.,  {Lithwick} Y.,  {Perets} H.~B.,   {Portegies
  Zwart} S.~F.,  2017, \mn@doi [\mnras] {10.1093/mnras/stw2370}, \href
  {https://ui.adsabs.harvard.edu/abs/2017MNRAS.464..688H} {464, 688}

\bibitem[\protect\citeauthoryear{{Holman}, {Touma}  \& {Tremaine}}{{Holman}
  et~al.}{1997}]{HTT1997}
{Holman} M.,  {Touma} J.,   {Tremaine} S.,  1997, \mn@doi [\nat]
  {10.1038/386254a0}, \href
  {https://ui.adsabs.harvard.edu/abs/1997Natur.386..254H} {386, 254}

\bibitem[\protect\citeauthoryear{{Hunter}}{{Hunter}}{2007}]{Hunter2007}
{Hunter} J.~D.,  2007, \mn@doi [Computing in Science and Engineering]
  {10.1109/MCSE.2007.55}, \href
  {https://ui.adsabs.harvard.edu/abs/2007CSE.....9...90H} {9, 90}

\bibitem[\protect\citeauthoryear{{Hut}}{{Hut}}{1981}]{Hut1981}
{Hut} P.,  1981, \aap, \href
  {https://ui.adsabs.harvard.edu/abs/1981A&A....99..126H} {99, 126}

\bibitem[\protect\citeauthoryear{{Jura}}{{Jura}}{2003}]{Jura2003}
{Jura} M.,  2003, \mn@doi [\apjl] {10.1086/374036}, \href
  {https://ui.adsabs.harvard.edu/abs/2003ApJ...584L..91J} {584, L91}

\bibitem[\protect\citeauthoryear{{Kalirai}, {Hansen}, {Kelson}, {Reitzel},
  {Rich}  \& {Richer}}{{Kalirai} et~al.}{2008}]{Kalirai+2008}
{Kalirai} J.~S.,  {Hansen} B. M.~S.,  {Kelson} D.~D.,  {Reitzel} D.~B.,  {Rich}
  R.~M.,   {Richer} H.~B.,  2008, \mn@doi [\apj] {10.1086/527028}, \href
  {https://ui.adsabs.harvard.edu/abs/2008ApJ...676..594K} {676, 594}

\bibitem[\protect\citeauthoryear{{Koester}, {G{\"a}nsicke}  \&
  {Farihi}}{{Koester} et~al.}{2014}]{KGF2014}
{Koester} D.,  {G{\"a}nsicke} B.~T.,   {Farihi} J.,  2014, \mn@doi [\aap]
  {10.1051/0004-6361/201423691}, \href
  {https://ui.adsabs.harvard.edu/abs/2014A&A...566A..34K} {566, A34}

\bibitem[\protect\citeauthoryear{{Kozai}}{{Kozai}}{1962}]{Kozai1962}
{Kozai} Y.,  1962, \mn@doi [\aj] {10.1086/108790}, \href
  {https://ui.adsabs.harvard.edu/abs/1962AJ.....67..591K} {67, 591}

\bibitem[\protect\citeauthoryear{{Lagos}, {Schreiber}, {Zorotovic},
  {G{\"a}nsicke}, {Ronco}  \& {Hamers}}{{Lagos} et~al.}{2021}]{Lagos+2020}
{Lagos} F.,  {Schreiber} M.~R.,  {Zorotovic} M.,  {G{\"a}nsicke} B.~T.,
  {Ronco} M.~P.,   {Hamers} A.~S.,  2021, \mn@doi [\mnras]
  {10.1093/mnras/staa3703}, \href
  {https://ui.adsabs.harvard.edu/abs/2021MNRAS.501..676L} {501, 676}

\bibitem[\protect\citeauthoryear{{Li}, {Lai}, {Anderson}  \& {Pu}}{{Li}
  et~al.}{2021}]{Li+2020}
{Li} J.,  {Lai} D.,  {Anderson} K.~R.,   {Pu} B.,  2021, \mn@doi [\mnras]
  {10.1093/mnras/staa3779}, \href
  {https://ui.adsabs.harvard.edu/abs/2021MNRAS.501.1621L} {501, 1621}

\bibitem[\protect\citeauthoryear{{Lidov}}{{Lidov}}{1962}]{Lidov1962}
{Lidov} M.~L.,  1962, \mn@doi [\planss] {10.1016/0032-0633(62)90129-0}, \href
  {https://ui.adsabs.harvard.edu/abs/1962P&SS....9..719L} {9, 719}

\bibitem[\protect\citeauthoryear{{Liu} \& {Lai}}{{Liu} \& {Lai}}{2019}]{LL2019}
{Liu} B.,  {Lai} D.,  2019, \mn@doi [\mnras] {10.1093/mnras/sty3432}, \href
  {https://ui.adsabs.harvard.edu/abs/2019MNRAS.483.4060L} {483, 4060}

\bibitem[\protect\citeauthoryear{{Liu}, {Mu{\~n}oz}  \& {Lai}}{{Liu}
  et~al.}{2015}]{LML2015}
{Liu} B.,  {Mu{\~n}oz} D.~J.,   {Lai} D.,  2015, \mn@doi [\mnras]
  {10.1093/mnras/stu2396}, \href
  {https://ui.adsabs.harvard.edu/abs/2015MNRAS.447..747L} {447, 747}

\bibitem[\protect\citeauthoryear{{Livio} \& {Soker}}{{Livio} \&
  {Soker}}{1984}]{LS1984}
{Livio} M.,  {Soker} N.,  1984, \mn@doi [\mnras] {10.1093/mnras/208.4.763},
  \href {https://ui.adsabs.harvard.edu/abs/1984MNRAS.208..763L} {208, 763}

\bibitem[\protect\citeauthoryear{{Maldonado}, {Villaver}, {Mustill},
  {Ch{\'a}vez}  \& {Bertone}}{{Maldonado} et~al.}{2021}]{Maldonado+2020}
{Maldonado} R.~F.,  {Villaver} E.,  {Mustill} A.~J.,  {Ch{\'a}vez} M.,
  {Bertone} E.,  2021, \mn@doi [\mnras] {10.1093/mnrasl/slaa193}, \href
  {https://ui.adsabs.harvard.edu/abs/2021MNRAS.501L..43M} {501, L43}

\bibitem[\protect\citeauthoryear{{Mu{\~n}oz} \& {Petrovich}}{{Mu{\~n}oz} \&
  {Petrovich}}{2020}]{MP2020}
{Mu{\~n}oz} D.~J.,  {Petrovich} C.,  2020, \mn@doi [\apjl]
  {10.3847/2041-8213/abc564}, \href
  {https://ui.adsabs.harvard.edu/abs/2020ApJ...904L...3M} {904, L3}

\bibitem[\protect\citeauthoryear{{Mu{\~n}oz}, {Lai}  \& {Liu}}{{Mu{\~n}oz}
  et~al.}{2016}]{MLL2016}
{Mu{\~n}oz} D.~J.,  {Lai} D.,   {Liu} B.,  2016, \mn@doi [\mnras]
  {10.1093/mnras/stw983}, \href
  {https://ui.adsabs.harvard.edu/abs/2016MNRAS.460.1086M} {460, 1086}

\bibitem[\protect\citeauthoryear{{Mustill} \& {Villaver}}{{Mustill} \&
  {Villaver}}{2012}]{MV2012}
{Mustill} A.~J.,  {Villaver} E.,  2012, \mn@doi [\apj]
  {10.1088/0004-637X/761/2/121}, \href
  {https://ui.adsabs.harvard.edu/abs/2012ApJ...761..121M} {761, 121}

\bibitem[\protect\citeauthoryear{{Mustill}, {Veras}  \& {Villaver}}{{Mustill}
  et~al.}{2014}]{MVV2014}
{Mustill} A.~J.,  {Veras} D.,   {Villaver} E.,  2014, \mn@doi [\mnras]
  {10.1093/mnras/stt1973}, \href
  {https://ui.adsabs.harvard.edu/abs/2014MNRAS.437.1404M} {437, 1404}

\bibitem[\protect\citeauthoryear{{Mustill}, {Villaver}, {Veras}, {G{\"a}nsicke}
   \& {Bonsor}}{{Mustill} et~al.}{2018}]{Mustill+2018}
{Mustill} A.~J.,  {Villaver} E.,  {Veras} D.,  {G{\"a}nsicke} B.~T.,   {Bonsor}
  A.,  2018, \mn@doi [\mnras] {10.1093/mnras/sty446}, \href
  {https://ui.adsabs.harvard.edu/abs/2018MNRAS.476.3939M} {476, 3939}

\bibitem[\protect\citeauthoryear{{Naoz}, {Farr}  \& {Rasio}}{{Naoz}
  et~al.}{2012}]{NFR2012}
{Naoz} S.,  {Farr} W.~M.,   {Rasio} F.~A.,  2012, \mn@doi [\apjl]
  {10.1088/2041-8205/754/2/L36}, \href
  {https://ui.adsabs.harvard.edu/abs/2012ApJ...754L..36N} {754, L36}

\bibitem[\protect\citeauthoryear{{Pejcha}, {Antognini}, {Shappee}  \&
  {Thompson}}{{Pejcha} et~al.}{2013}]{Pejcha+2013}
{Pejcha} O.,  {Antognini} J.~M.,  {Shappee} B.~J.,   {Thompson} T.~A.,  2013,
  \mn@doi [\mnras] {10.1093/mnras/stt1281}, \href
  {https://ui.adsabs.harvard.edu/abs/2013MNRAS.435..943P} {435, 943}

\bibitem[\protect\citeauthoryear{{Petrovich}}{{Petrovich}}{2015a}]{Petrovich2015}
{Petrovich} C.,  2015a, \mn@doi [\apj] {10.1088/0004-637X/799/1/27}, \href
  {https://ui.adsabs.harvard.edu/abs/2015ApJ...799...27P} {799, 27}

\bibitem[\protect\citeauthoryear{{Petrovich}}{{Petrovich}}{2015b}]{Petrovich2015b}
{Petrovich} C.,  2015b, \mn@doi [\apj] {10.1088/0004-637X/805/1/75}, \href
  {https://ui.adsabs.harvard.edu/abs/2015ApJ...805...75P} {805, 75}

\bibitem[\protect\citeauthoryear{{Petrovich} \& {Mu{\~n}oz}}{{Petrovich} \&
  {Mu{\~n}oz}}{2017}]{PM2017}
{Petrovich} C.,  {Mu{\~n}oz} D.~J.,  2017, \mn@doi [\apj]
  {10.3847/1538-4357/834/2/116}, \href
  {https://ui.adsabs.harvard.edu/abs/2017ApJ...834..116P} {834, 116}

\bibitem[\protect\citeauthoryear{{Pichierri}, {Morbidelli}  \&
  {Lai}}{{Pichierri} et~al.}{2017}]{PML2017}
{Pichierri} G.,  {Morbidelli} A.,   {Lai} D.,  2017, \mn@doi [\aap]
  {10.1051/0004-6361/201730936}, \href
  {https://ui.adsabs.harvard.edu/abs/2017A&A...605A..23P} {605, A23}

\bibitem[\protect\citeauthoryear{{Stephan}, {Naoz}  \& {Zuckerman}}{{Stephan}
  et~al.}{2017}]{SNZ2017}
{Stephan} A.~P.,  {Naoz} S.,   {Zuckerman} B.,  2017, \mn@doi [\apjl]
  {10.3847/2041-8213/aa7cf3}, \href
  {https://ui.adsabs.harvard.edu/abs/2017ApJ...844L..16S} {844, L16}

\bibitem[\protect\citeauthoryear{{Stephan}, {Naoz}  \& {Gaudi}}{{Stephan}
  et~al.}{2020}]{SNG2020}
{Stephan} A.~P.,  {Naoz} S.,   {Gaudi} B.~S.,  2020, arXiv e-prints, \href
  {https://ui.adsabs.harvard.edu/abs/2020arXiv201010534S} {p. arXiv:2010.10534}

\bibitem[\protect\citeauthoryear{{Teyssandier}, {Lai}  \& {Vick}}{{Teyssandier}
  et~al.}{2019}]{TLV2019}
{Teyssandier} J.,  {Lai} D.,   {Vick} M.,  2019, \mn@doi [\mnras]
  {10.1093/mnras/stz1011}, \href
  {https://ui.adsabs.harvard.edu/abs/2019MNRAS.486.2265T} {486, 2265}

\bibitem[\protect\citeauthoryear{{Vanderburg} et~al.,}{{Vanderburg}
  et~al.}{2015}]{Vanderburg+2015}
{Vanderburg} A.,  et~al., 2015, \mn@doi [\nat] {10.1038/nature15527}, \href
  {https://ui.adsabs.harvard.edu/abs/2015Natur.526..546V} {526, 546}

\bibitem[\protect\citeauthoryear{{Vanderburg} et~al.,}{{Vanderburg}
  et~al.}{2020}]{Vanderburg+2020}
{Vanderburg} A.,  et~al., 2020, \mn@doi [\nat] {10.1038/s41586-020-2713-y},
  \href {https://ui.adsabs.harvard.edu/abs/2020arXiv200907282V} {585, 363}

\bibitem[\protect\citeauthoryear{{Veras} \& {Fuller}}{{Veras} \&
  {Fuller}}{2019}]{VF2019}
{Veras} D.,  {Fuller} J.,  2019, \mn@doi [\mnras] {10.1093/mnras/stz2339},
  \href {https://ui.adsabs.harvard.edu/abs/2019MNRAS.489.2941V} {489, 2941}

\bibitem[\protect\citeauthoryear{{Veras} \& {G{\"a}nsicke}}{{Veras} \&
  {G{\"a}nsicke}}{2015}]{VG2015}
{Veras} D.,  {G{\"a}nsicke} B.~T.,  2015, \mn@doi [\mnras]
  {10.1093/mnras/stu2475}, \href
  {https://ui.adsabs.harvard.edu/abs/2015MNRAS.447.1049V} {447, 1049}

\bibitem[\protect\citeauthoryear{{Veras}, {Mustill}, {G{\"a}nsicke},
  {Redfield}, {Georgakarakos}, {Bowler}  \& {Lloyd}}{{Veras}
  et~al.}{2016}]{Veras+2016}
{Veras} D.,  {Mustill} A.~J.,  {G{\"a}nsicke} B.~T.,  {Redfield} S.,
  {Georgakarakos} N.,  {Bowler} A.~B.,   {Lloyd} M. J.~S.,  2016, \mn@doi
  [\mnras] {10.1093/mnras/stw476}, \href
  {https://ui.adsabs.harvard.edu/abs/2016MNRAS.458.3942V} {458, 3942}

\bibitem[\protect\citeauthoryear{{Vick} \& {Lai}}{{Vick} \&
  {Lai}}{2018}]{VL2018}
{Vick} M.,  {Lai} D.,  2018, \mn@doi [\mnras] {10.1093/mnras/sty225}, \href
  {https://ui.adsabs.harvard.edu/abs/2018MNRAS.476..482V} {476, 482}

\bibitem[\protect\citeauthoryear{{Vick}, {Lai}  \& {Anderson}}{{Vick}
  et~al.}{2019}]{VLA2019}
{Vick} M.,  {Lai} D.,   {Anderson} K.~R.,  2019, \mn@doi [\mnras]
  {10.1093/mnras/stz354}, \href
  {https://ui.adsabs.harvard.edu/abs/2019MNRAS.484.5645V} {484, 5645}

\bibitem[\protect\citeauthoryear{{Vigan} et~al.,}{{Vigan}
  et~al.}{2017}]{Vigan+2017}
{Vigan} A.,  et~al., 2017, \mn@doi [\aap] {10.1051/0004-6361/201630133}, \href
  {https://ui.adsabs.harvard.edu/abs/2017A&A...603A...3V} {603, A3}

\bibitem[\protect\citeauthoryear{{Villaver} \& {Livio}}{{Villaver} \&
  {Livio}}{2007}]{VL2007}
{Villaver} E.,  {Livio} M.,  2007, \mn@doi [\apj] {10.1086/516746}, \href
  {https://ui.adsabs.harvard.edu/abs/2007ApJ...661.1192V} {661, 1192}

\bibitem[\protect\citeauthoryear{{Virtanen} et~al.,}{{Virtanen}
  et~al.}{2020}]{Virtanen+2020}
{Virtanen} P.,  et~al., 2020, \mn@doi [Nature Methods]
  {10.1038/s41592-019-0686-2}, \href
  {https://ui.adsabs.harvard.edu/abs/2020NatMe..17..261V} {17, 261}

\bibitem[\protect\citeauthoryear{{Wilson}, {Farihi}, {G{\"a}nsicke}  \&
  {Swan}}{{Wilson} et~al.}{2019}]{Wilson+2019}
{Wilson} T.~G.,  {Farihi} J.,  {G{\"a}nsicke} B.~T.,   {Swan} A.,  2019,
  \mn@doi [\mnras] {10.1093/mnras/stz1050}, \href
  {https://ui.adsabs.harvard.edu/abs/2019MNRAS.487..133W} {487, 133}

\bibitem[\protect\citeauthoryear{{Wu}}{{Wu}}{2018}]{Wu2018}
{Wu} Y.,  2018, \mn@doi [\aj] {10.3847/1538-3881/aaa970}, \href
  {https://ui.adsabs.harvard.edu/abs/2018AJ....155..118W} {155, 118}

\bibitem[\protect\citeauthoryear{{Wu} \& {Lithwick}}{{Wu} \&
  {Lithwick}}{2011}]{WL2011}
{Wu} Y.,  {Lithwick} Y.,  2011, \mn@doi [\apj] {10.1088/0004-637X/735/2/109},
  \href {https://ui.adsabs.harvard.edu/abs/2011ApJ...735..109W} {735, 109}

\bibitem[\protect\citeauthoryear{{Wu} \& {Murray}}{{Wu} \&
  {Murray}}{2003}]{WM2003}
{Wu} Y.,  {Murray} N.,  2003, \mn@doi [\apj] {10.1086/374598}, \href
  {https://ui.adsabs.harvard.edu/abs/2003ApJ...589..605W} {589, 605}

\bibitem[\protect\citeauthoryear{{Zuckerman}}{{Zuckerman}}{2014}]{Zuckerman2014}
{Zuckerman} B.,  2014, \mn@doi [\apjl] {10.1088/2041-8205/791/2/L27}, \href
  {https://ui.adsabs.harvard.edu/abs/2014ApJ...791L..27Z} {791, L27}

\bibitem[\protect\citeauthoryear{{Zuckerman}, {Koester}, {Reid}  \&
  {H{\"u}nsch}}{{Zuckerman} et~al.}{2003}]{Zuckerman+2003}
{Zuckerman} B.,  {Koester} D.,  {Reid} I.~N.,   {H{\"u}nsch} M.,  2003, \mn@doi
  [\apj] {10.1086/377492}, \href
  {https://ui.adsabs.harvard.edu/abs/2003ApJ...596..477Z} {596, 477}

\bibitem[\protect\citeauthoryear{{Zuckerman}, {Melis}, {Klein}, {Koester}  \&
  {Jura}}{{Zuckerman} et~al.}{2010}]{Zuckerman+2010}
{Zuckerman} B.,  {Melis} C.,  {Klein} B.,  {Koester} D.,   {Jura} M.,  2010,
  \mn@doi [\apj] {10.1088/0004-637X/722/1/725}, \href
  {https://ui.adsabs.harvard.edu/abs/2010ApJ...722..725Z} {722, 725}

\bibitem[\protect\citeauthoryear{{van Sluijs} \& {Van Eylen}}{{van Sluijs} \&
  {Van Eylen}}{2018}]{vSvE2018}
{van Sluijs} L.,  {Van Eylen} V.,  2018, \mn@doi [\mnras]
  {10.1093/mnras/stx3068}, \href
  {https://ui.adsabs.harvard.edu/abs/2018MNRAS.474.4603V} {474, 4603}

\bibitem[\protect\citeauthoryear{{van der Walt}, {Colbert}  \&
  {Varoquaux}}{{van der Walt} et~al.}{2011}]{vDW+2011}
{van der Walt} S.,  {Colbert} S.~C.,   {Varoquaux} G.,  2011, \mn@doi
  [Computing in Science and Engineering] {10.1109/MCSE.2011.37}, \href
  {https://ui.adsabs.harvard.edu/abs/2011CSE....13b..22V} {13, 22}

\bibitem[\protect\citeauthoryear{{von Zeipel}}{{von
  Zeipel}}{1910}]{vonZeipel1910}
{von Zeipel} H.,  1910, \mn@doi [Astronomische Nachrichten]
  {10.1002/asna.19091832202}, \href
  {https://ui.adsabs.harvard.edu/abs/1910AN....183..345V} {183, 345}

\makeatother
\end{thebibliography}
\bibliographystyle{mnras.bst}

\label{lastpage}

\end{document}